\documentstyle[12pt,leqno]{article}

\begin{document}
\baselineskip=1.\baselineskip

\makebox[7cm]

\vskip 2cm

\begin{center}
{\Large \bf On QCD  $Q^2$-evolution\\[0.2cm]
 of Deuteron Structure Function $F_2^D(x_D,Q^2)$\\[0.2cm]
  for $x_D>1$}\\
\end{center}
\vskip 2cm

\begin{center}
{\bf Alexander V. Sidorov}\\[0.3cm]
{\it Bogoliubov Laboratory of Theoretical Physics,\\
 Joint Institute for Nuclear Research,\\
141980 Dubna, Moscow Region, Russia}

\vskip 0.5cm

{\bf Michael V.Tokarev}\\[0.3cm]
{\it Laboratory of High Energies, \\
Joint Institute for Nuclear Research, \\
141980, Dubna, Moscow Region, Russia}
\end{center}

\vskip 1.cm
{\bf{Summary.$-$}\it{
The deep-inelastic deuteron structure function (SF) $F_2^D(x_D,Q^2)$
in the covariant approach in light-cone variables is considered.
The $x_D$ and $Q^2$-dependences of SF are calculated.
The QCD analysis  of generated data  both for non-cumulative $x_D<1$
and cumulative  $x_D>1$  ranges  was performed.
It was shown that $Q^2$-evolution of SF is valid  for ranges
$0.275<x_D<0.85$ and  $1.1<x_D<1.4$ for the same value of
QCD scale parameter  ${\Lambda}$.
It was found the $x_D$-dependence
 of SF for the ranges is essentially different.}} \\[5mm]

{\large {\bf1.$-$ Introduction}}\\

The progress of perturbative Quantum Chromodynamics (QCD) in the description
of the high energy physics of strong interactions is considerable
\cite{Altarelli}.
Deep-inelastic scattering (DIS) of leptons provides the data on nucleon
structure functions (SF) which could be used for the most precise comparison
with QCD. The accuracy of SF data is sufficient to be sensitive to nuclear
effects in DIS, e.g. the EMC effect \cite{Arneodo}.
 A direct evidence of the nucleus structure
in DIS has been obtained by measuring the carbon SF \cite{Carbon}
in the kinematical region  forbidden for scattering on a free
 nucleon (the Bjorken kinematical variable $x=Q^2/2(pq)$ is more than
1 in the range). The region is known as a cumulative region
 \cite{Baldin71,Stavin79}
 and it is intensively
investigated both experimentally and theoretically in proton-nucleus
 and nucleus-nucleus collisions \cite{Stavin79,Baldin84}.
 An important matter of the
relativistic nuclear physics is
to adduce arguments for applying QCD to
analyse SF at $x_D>1$. The deuteron as a simplest nuclear system is an
excellent object to investigate the problem of relativistic theory of a
deuteron.

Dependence of the deuteron structure functions $F_2^D$ on $x_D$
in the cumulative range has been studied by many authors.
Different models
(the few-nucleon correlation model  \cite{Frank},
the flucton model  \cite{Titov} ) are used to
describe a high momentum component of the deuteron wave function.
The microscopic picture for
cumulative phenomena on the deuteron near the threshold  based on
perturbative QCD has been developed in \cite{Vechernin}.

The dependence of the nuclear structure function ratio
$R^{C/D}=F_2^C/F_2^D$ on $x_D$ has been considered in \cite{TMV95}.
In the analysis the BCDMS experimental data on $F_2^C$
\cite{Carbon} have been used.  The deuteron SF has been
 calculated in the light-cone  covariant approach \cite{Tok91} and
it was found that the ratio $R^{C/D}(x_D)$ reveals an exponent
growth $\sim exp(\alpha \cdot x_D)$ for $x_D>1$  with the slope parameter
$\alpha \simeq 6.6$ and differs
significantly from the ratio behaviour for $x_D<1$. The latter is
in good agreement with experimental data \cite{BCDMS89,NMC91,SLAC92}.
It is considered  that the ratio
$R^{C/D}$ for $x_D<1$
describes quark distributions in a nucleon of a nucleus. The ratio
$R^{C/D}$ for $x_D>1$  describes  quark distributions in a nucleus.
In \cite{Averich} similar exponent  dependence of
the  inclusive pion backward     cross section ratio
 $R^{A/D} \sim {\rho }_{A}^{\pi} (X)/ {\rho}_D^{\pi}(X)$ on
 the cumulative number $X$ \cite{Stavin76} for the
$p+A \rightarrow {\pi}(180^o)+ ...$  process was found.
It was concluded that this dependence demonstrates the change
of the regime from hadron to quark  degrees of freedom.
We can consider that  the exponent growth of the similar ratio $R^{A/D}$
for different processes (deep-inelastic lepton scattering,
deep-inelastic nucler reactions) in the cumulative range
is the general feature of quark interactions in the superdensity
matter.

An important matter in the description of DIS is whether
  or not the $Q^2$-evolution
of the structure function is possible in the framework of QCD in the cumulative range.
If it is possible, the
QCD analysis could  be used to study the transition regime
 and to determine quark distribution functions in nuclei.

In the present paper, the QCD analysis of data on the deep-inelastic deuteron
structure function $F_2^D(x_D,Q^2)$ for $x_D<1$ and $x_D>1$ is performed.
The data on the $x_D$ and $Q^2$-dependence of SF  are simulated by
the relativistic deuteron model \cite{Tok91}.
The simulated data for  $F_2^D(x_D,Q^2)$  for the range
$x_D<1$ are in good agreement with experimental data
\cite{BCDMS89}. The experimental
data  for the deuteron  SF  in the range $x_D>1$ for high $Q^2$ are absent.
These data would be very interesting both to verify deuteron models
 and methods of describing  relativistic nuclear systems and to study
the high momentum component of the relativistic deuteron wave function (RDWF).
The QCD analysis of data  is based on
the expansion method of SF over Jacobi polynomials
\cite{Jacobi}-\cite{Kriv}.
The $Q^2$-evolution of the moments of the structure function $F_2^D$
is found as the solution of the corresponding renormalization
equation.  Effectivity of the method to control higher
perturbative QCD corrections and to investigate sensitivity  of the
QCD scale parameter $\Lambda$ were shown in \cite{Kriv}.
It was found that the data on $F_2^D(x_D,Q^2)$ obey the
$Q^2$-evolution of  SF in the framework of QCD both
for $0.275<x_D<0.85$ and  $1.1<x_D<1.4$  ranges in accordance with the
criterium
${\chi}^2$ very well. The value of the parameter ${\Lambda}$
is determined to be the same for both the regions. We would like to note
that the parameters of structure function parametrisations differ
significantly
for these ranges and this gives us some evidence for two regimes
of the $x_D$-dependence of the structure function $F_2^D(x_D,Q^2)$. \\

{\large {\bf 2.$-$ Deuteron  Structure Function  $F^D_2(x_D,Q^2)$}}\\

The deuteron structure functions  $W_{1,2}(\nu, Q^2)$  are related to
 the imaginary part
 of the forward scattering amplitude  of a virtual photon on a deuteron
$W_{\mu \nu}$ by the standard formula

\begin{equation}
  W_{\mu \nu}^D = -(g_{\mu \nu} - q_{\mu}q_{\nu}/q^2)\cdot W_1^D
  + (p_{\mu}-q_{\mu}(pq)/q^2)(p_{\nu}-q_{\nu}(pq)/q^2)\cdot W_2^D /M^2.
\end{equation}
Here $q, p$ are momenta of a photon and a deuteron; $M$ is the deuteron mass.

In the relativistic impulse approximation (RIA) the  forward scattering
amplitude of
the virtual photon $\gamma^{\ast}$  on the
deuteron $A_{\mu \nu}^D$  is defined
via a similar scattering amplitude on the
nucleon $A_{\mu \nu}^N$  as follows
\begin{equation}
A_{\mu \nu}^D (q,p) = \int \frac{d^4 k_1}{(2\pi)^4i} Sp \{A_{\mu \nu}^N(q,k_1)
\cdot T(s_1,k_1)\}.
\end{equation}
In the expression (2) $T(s_1,k_1)$ is the amplitude of the forward
$\bar N-D $ scattering and  the conventional notations
$Q^2 = -q^2 > 0,\ \nu \equiv (pq),\ s_1 = (p - k_1)^2$ are  used.
The integration is carried out over the
 active  nucleon momentum $k_1$.
 The calculation of the imaginary part of the amplitude $A_{\mu \nu}^D$
 gives us the  possibility
to put the nucleon spectator with the momentum $k = p - k_1$ on mass shell.
Therefore the tensor $W_{\mu \nu}^D$ is expressed via the DNN vertex with one
 nucleon on mass shell.
The vertex is described by the function $\Gamma_{\alpha} (k_1)$
and  depends on one variable $k_1$. The vector index $\alpha$
 characterizes the deuteron spin.
With the relation between the RDWF and the vertex function
 $\Gamma_{\alpha} (k_1)$:
$\psi_{\alpha} (k_1) = \Gamma_{\alpha} (k_1)\cdot (m + \hat k_1)^{-1},$
the expression for the tensor $W_{\mu \nu}^{\alpha \beta}$  can be written as

\begin{equation}
W_{\mu \nu}^{D} = \int \frac{d^4 k}{(2\pi)^3}
\delta (m^2-k^2) {\Theta (k_0)} {\Theta (p_{+}-k_{+})}\ Sp \{w_{\mu \nu}^N
\cdot \bar {\psi}^{\alpha}(k_1)\cdot (m+\hat k)\cdot
\psi^{\beta}(k_1)\}\cdot {\rho}_{\alpha \beta}(p).
\end{equation}
Here ${\rho}_{\alpha \beta}(p)
=-(g_{\alpha \beta}-p_{\alpha}p_{\alpha}/M^2)/3$ is
 the deuteron polarization density matrix and the
 $\Theta $ - function and light-cone variables
$(k_{\pm} = k_{0} \pm k_{3}, k_{\bot})$ are used.
The vertex function $\Gamma_{\alpha} (k_1)$  is defined via 4 scalar functions
$a_i (k_1^2)\ (i=1-4)$ and  has  the form

\begin{equation}
\Gamma_{\alpha} (k_1) = k_{1 \alpha}[a_1(k_1^2)+a_2(k_1^2)(m+\hat k_1)]
+ \gamma_{\alpha} [a_3(k_1^2)+a_4(k_1^2)(m+\hat k_1)].
\end{equation}
The  relativization  procedure  of deuteron wave function
${\psi}_{\alpha}$  has been proposed in \cite{Tok91,Tok83}.
The scalar functions $a_{i}(k_1^2)$ have been constructed
 in the form
of a sum of pole terms. Some pole positions and residues have been found
 from the comparison  of our RDWF
in the nonrelativistic limit             with the known
 nonrelativistic deuteron wave function.
For the latter  the Paris wave function \cite{Paris} was taken.
Other parameters  were fixed from the description of the static
characteristics of the deuteron
(an electric charge - $G_e(0)=1(e)$, magnetic - $G_m(0)=\mu_D(e/2M_D)$
and quadrupole - $G_Q(0)=Q_D(e/M_D^2)$ moments) in the relativistic
impulse approximation.

The calculation of (3) in the light-cone variables
gives the final expression for the deuteron SF  $F_2^D \equiv \nu W_2^D$

\begin{equation}
F_2^D (x_D, Q^2) = \int_{z}^{1} dx {\ }{d^2}k_{\bot}\ p(x,k_{\bot})
\cdot  F_2^N(z /x, Q^2),
\end{equation}
where $x_D=2z,\ 0<z<1$.
The nucleon SF $F_2^N=(F_2^p+F_2^n)/2$ is defined by the proton and
neutron ones.
The positive function $p(x,k_{\bot})$ describes the probability that
 the active nucleon carries away the fraction of the deuteron momentum
 $x = k_{1+}/p_{+}$
 and the transverse momentum $k_{\bot}$ in the infinite momentum frame.
 It is expressed via  the vertex function  $\Gamma_{\alpha} (k_1)$
 as follows

\begin{equation}
p(x,k_{\bot}) \propto  Sp\{ \bar {\psi}^{\alpha}(k_1)\cdot (m+\hat k)\cdot
\psi^{\beta}(k_1)\cdot \hat q
 \cdot\rho_{\alpha \beta}(p)
  \}.
\end{equation}

Note that in the approach used the distribution function
$p(x,k_{\bot})$ includes not only the usual $S$- and $D$-wave components
of the deuteron but the $P$-component too. The latter describes
the contribution of the $N\bar N$-pair production.

In the RIA the deuteron SF $F_2^D$ is defined by (5) as a sum
of the proton and neutron SF integrated on the $x$ and $k_{\bot}$.
 The NMC data \cite{NMC91} on the ratio
 $R_F^{D/p}=F_2^D / F_2^p$ and  $F_2^p$  and the relativistic deuteron
model have been used  to extract the neutron SF  $F_2^n$.
For the latter the parametrisation

\begin{equation}
F_2^n (x,Q^2)= (1-0.75x)(1-0.15 {\sqrt x} (1-x))
\cdot F_2^p (x,Q^2)
\end{equation}
has been obtained in \cite{Tok94}.
The results  for the absolute value
 of  $F_2^D$ calculated within the parametrisation of the neutron SF
   are in good agreement with experimental data both in
low  \cite{NMC91} and high  \cite{BCDMS89,SLAC92}  $x_D$-ranges. \\

{\large {\bf 3.$-$ Proton Structure Function  $F^p_2(x,Q^2)$}} \\

In our analysis we
 shall use the parametrisation of $F^p_2(x,Q^2)$ given in
 \cite{NMC91}.  The parametrization describes the NMC, SLAC and BCDMS
data very well. The verification of this fit in the region
 $0.006<x< 0.6$  gives ${\Lambda}=200\ MeV$.
 To parametrize $F^p_2(x,Q^2)$ at $0.55<x< 1$, we have made the
 QCD fit of parametrisation
\cite{NMC91} considering it as an "experimental points" at
$0.275< x< 0.55$ with the leading order nonsinglet evolution of the
moments of $F^p_2(x,Q^2)$ with
$\Lambda=200\ MeV$. The SF at the fixed
 point $Q^2_0=10\ (GeV/c)^2$ was parametrized as

\begin{equation}
  F^p_{2}(x,Q_0^2)=Ax^{B}(1-x)^{C}~(1+\gamma x).
\label{e10}
\end{equation}

The parameters A, B, C and $\gamma$  in Eq. (\ref{e10})
are free parameters and  are determined by the fit of the data.
Then, on the basis of expression (\ref{e10})  the values of SF for
$0.55< x < 1$ were calculated.
To achieve agreement  between QCD - evolution results
and parametrization \cite{NMC91}, the former should be multiplied by
 the factor $R(x,Q^2)$ for $0.55<x< 1$:

\begin{equation}
  R(x,Q^2)=(1.+17.611(x-0.55)^{-(3.+0.661log(Q^2/Q^2_0))} )\Theta(x-0.55).
\label{NMC91}
\end{equation}

The modified parametrization of the proton SF in a wide range
of the Bjorken variable $x$ is used  as an "experimental"
input for the deuteron model described  above to simulate
the data on the deuteron SF.  The data for the deuteron SF
are simulated for
$x_D=0.1-1.8$ and $Q^2=17-230\ (GeV/c)^2$.

Figure 1 shows  the $x_D$-dependence of $F_2^D(x_D,Q^2)$
 at $Q^2=61.5\ (GeV/c)^2$. One can see that the SF falls down drastically
for   $x_D>1$. The open points are BCDMS data for SF
$F_2^C$ for the $\mu + ^{12}C\rightarrow {\mu}^{'}+...$ process \cite{Carbon}.
The model-simulated data for $F_2^D(x_D,Q^2)$
are in a qualitative agreement with BCDMS data.
We would like  to note that the more detail  calculations \cite{TMV95}
of the dependence  of the nuclear structure function ratio
$R^{C/D}=F_2^C/F_2^D$ on $x_D$ with
 RDWF \cite{Tok91} have shown that the ratio
is similar to the experimental data  for $0.01<x_D<1$
\cite{ratio} and exponentially growthes for $x_D>1$.             \\

{\large {\bf 4.$-$ Method and Results of Analysis}}           \\

Now we can apply the nonsinglet  QCD fit to the data on the deuteron SF
simulated in the previous section. We have reduced
the deuteron data to a standard
interval of the scaling variable $0<x<1$ by putting $x=x_D/2$
and have applied the method
of the QCD analysis based on the expansion of SF over the Jacobi polynomials
\cite{Jacobi}-\cite{Kriv}.
The SF is presented  as follows

\begin{equation}
F_2^{N_{max}} (x,Q^2)=x^{\alpha }{(1-x)}^{\beta } \sum_{n=0}^{N_{max}}
{\Theta }_n^{\a ,\beta } (x)
\sum_{j=0}^{n} c_{j}^{(n)} {(\alpha ,\beta )}
 M_{j+2} \left ( Q^{2}\right ),
\label{e7}
 n = 2,3, ...  \nonumber
\end{equation}

\begin{equation}
M_{n}(Q^2)
 =\left [ \frac{\alpha _{s}\left ( Q_{0}^{2}\right )}
{\alpha _{s}\left ( Q^{2}\right )}\right ]^{d_{n}}
 M_{n}(Q_{0}^2)
\label{e1}
\end{equation}
where  $d_{n}=-\gamma_{NS}^{0}/2\beta_0$,\
  $\beta_0=11-\frac{2}{3}N_f,\ N_f$ is the number of flavours,
$\gamma_{NS}^{(0)}$ is
the anomalous dimension in the one-loop perturbative QCD approximation:\\
$\gamma_{NS}^{(0)}=8/3\left[4S_1(n)-3-2/(n(n+1))\right]$, where
$S_1(n)=\sum_{j=1}^{n}1/j$.

The moments  $M_{n}(Q^2)$ of SF $F_2^D$ are defined by

\begin{equation}
M_{n}(Q^2) = \int_{0}^{1} dx\ x^{N-2}\ F_2^D(x,Q^2).
\end{equation}
The parametrization form of SF similar to (8) is used
at a fixed point $Q^2_0$:

\begin{equation}
  F^D_{2}(x,Q_0^2)=A{(x_D/2)}^{B}(1-x_D/2)^{C}~(1+\gamma x_D/2).
\label{FD}
\end{equation}
The constants $A$, $B$, $C$, $\gamma$
in (\ref{FD}) and the QCD scale parameter
$\Lambda$ are considered as free parameters to be determined
at $Q_0^2$.\\

The results of the fit are presented in Table 1.
We have found that values of $\Lambda$ are approximately
the same  in two different regions of $x_D$
and  not significantly  differ from the input proton parameter
 $\Lambda =200\ GeV$ \cite{NMC91}.
 The $x_D$-dependence  of SF for two ranges
$x_D=0.225-0.85$ and $x_D=1.1-1.4$~,  is determined separately.
It is established that the shapes of SF are essential different
(see Table 1). The obtained value  of $\chi^2_{d.f.}$
corresponds to 10\% for errors of "experimental" deuteron data.

\vskip 2cm

\begin{center}
{{\bf Table 1.} The results of the LO nonsinglet
QCD fit of the deuteron SF.  $N_f=4$, $17<Q^2<230\ (GeV/c)^2$,
 $N_{max}=12$. The errors of $F_2^D$ are put to be $10\%$.}
\vskip 1cm

\begin{tabular}{|c|l|l|}     \hline
                  & $0.275\leq x_D\leq 0.85$   &$1.1\leq x_D\leq 1.4$       \\  \hline
$\Lambda$ [MeV]   &     268   $\pm$     67     &     240   $\pm$     44     \\
$A$               &    8.39   $\pm$      1.50  & (6.26  $\pm 0.84)10^{-5}$  \\
$B$               &     0.898  $\pm$     0.065 &  -0.115    $\pm$ 0.015     \\
$C$               &    8.659   $\pm$      0.456&    11.07   $\pm$ 0.07      \\
$\gamma$          &  -2.034    $\pm$     0.055 &    3376.   $\pm$ 457.      \\
$\chi^2_{d.f.}$   &    12.0/121                &   34.4/68                  \\ \hline
\end{tabular}

\end{center}

Figure 2 shows the dependence of  $F_2^D(x_D,Q^2)$ on $Q^2$
for $x_D=1.1-1.8$. The results of the QCD fit  for $x_D=1.1-1.4$
are drawn by  solid lines, and the results of direct calculation
 by formula (5) are shown by open points.
Good agreement between the model-simulated data on $F_2^D(x_D,Q^2)$
 and results of the QCD fit is observed.

We would like to note that a more general task is a simultaneous
QCD  fit of $F_2^D$ in the whole range $x_D=0.-2.$ but it requires
 a rather complicated $x$-parametrization of SF instead
of (8). \\

{\large {\bf 5.$-$ Conclusion}} \\

 The QCD analysis of data on the deep-inelastic deuteron structure function
$F_2^D(x_D,Q^2)$ was performed.
 The data were simulated by the relativistic
deuteron model in the covariant approach in light-cone variables.
It was found that the data on $F_2^D(x_D,Q^2)$ obey the
$Q^2$-evolution of SF in the framework of QCD both
for $0.275<x_D<0.85$ and  $1.1<x_D<1.4$  in accordance with the criterium
${\chi}^2$ very well. The value of the parameter ${\Lambda}$
was determined to be the same for both the regions.
The parameters of structure function parametrizations were determined and
it was found that they differ  significantly
for these ranges. These results give some evidence
for two regimes of the $x_D$-dependence
of the structure function $F_2^D(x_D,Q^2)$~.

 We would like to emphasize that the QCD $Q^2$ - evolution of SF in the
cumulative range  ($x_D>1$) could be a crucial test for verification of both
the nuclear models and the parton model, and thus measurements of SF in DIS
for $x_D>1$ should be performed with a high accuracy
 at CERN, DESY and FERMILAB.
\vskip 1cm

{\Large \bf Acknowledgement}
\vskip 5mm

This research was partly supported by INTAS (International
Association for the Promotion of Cooperation with Scientists from the
Independent States of the Former Soviet Union) under Contract No. 93-1180
and by the Russian Foundation for Fundamental Research Grant No. 95-02-04314
and No. 95-02-05061. One of the authors (M.T.) thanks I.Zborovsky
for useful discussion on results and Institute Nuclear Physics
of Academy of Science of the Czech Republic  for hospitality during
the completion of the paper.


\newpage
Figure captions. \\[5mm]

Fig~1. ~Deep-inelastic  deuteron structure function  $F_2^D (x_D,Q^2)$.
Theoretical results have been obtained with the proton parametrization
$F_2^p$ taken from [13].
Experimental data for $F_2^C$
 are taken from  [3]: $\circ  - 61.5\ (GeV/c)^2$. \\[3mm]

Fig~2. ~Deep-inelastic  deuteron structure function  $F_2^D (x_D,Q^2)$.
Open points are generated "experimental"  data;
solid lines are results of the QCD fit of the data.

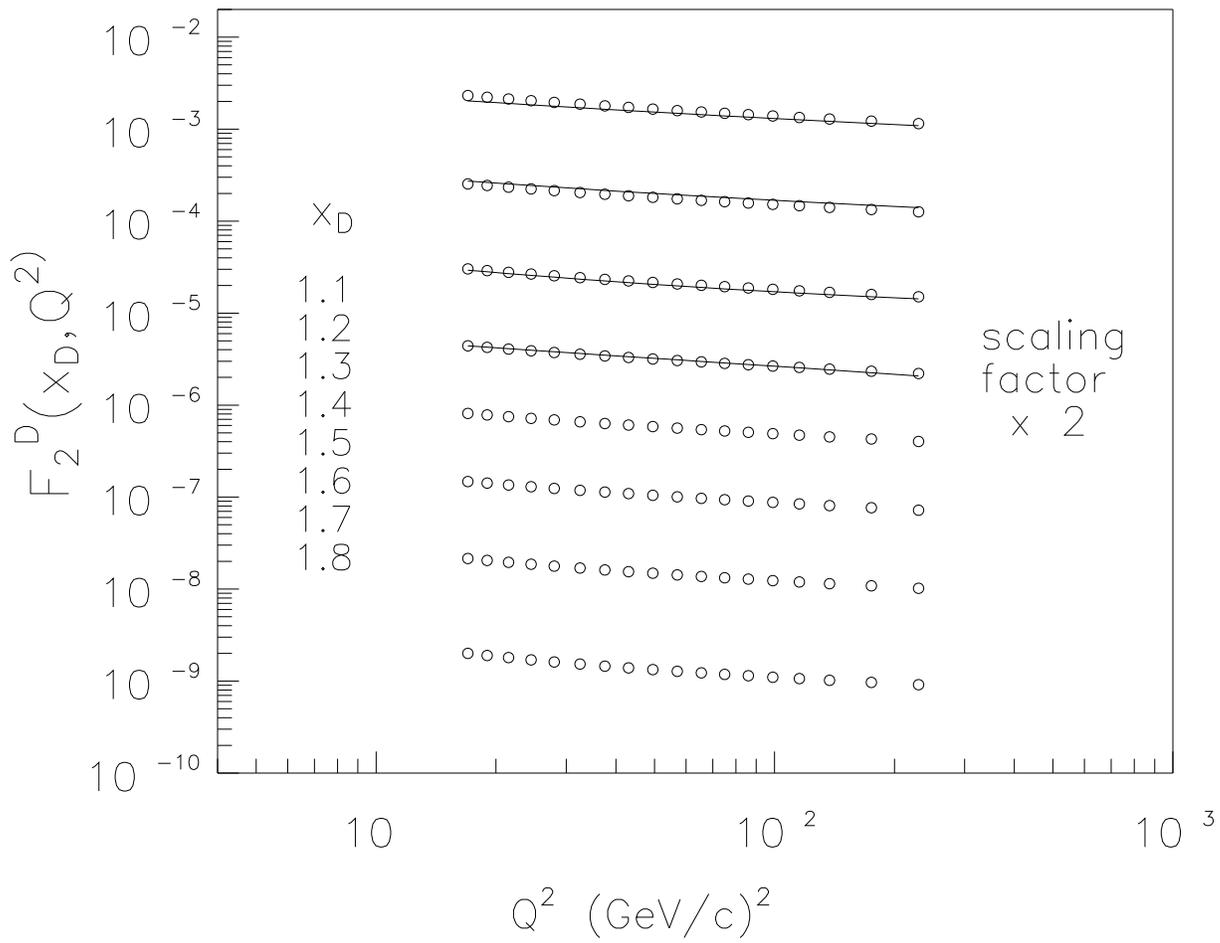
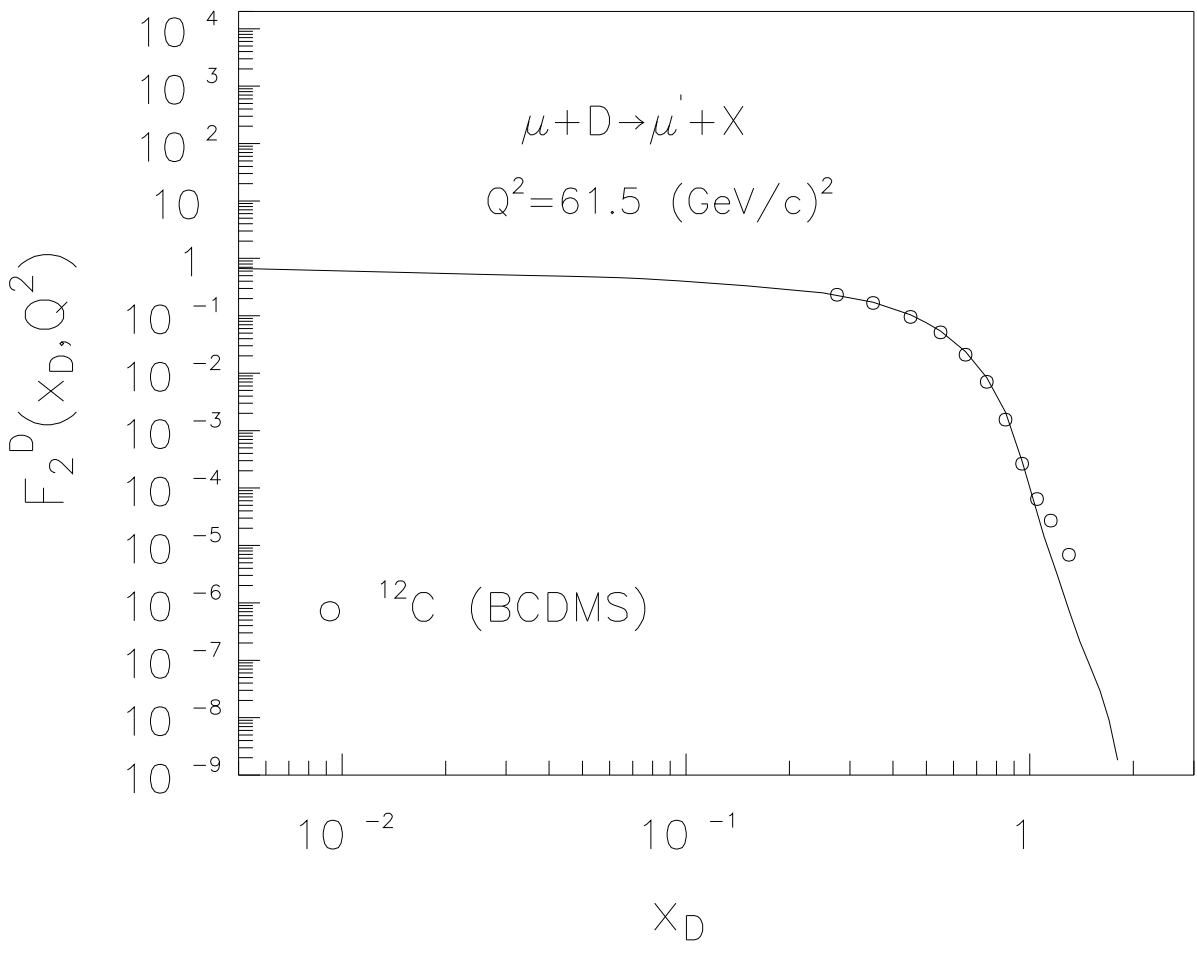
\end{document}